\documentstyle[preprint,aps]{revtex}
\begin{document}
\title{Static and dynamical properties of a magnetic impurity
in a strongly correlated electronic system}
\author{K. A. Hallberg}
\address{Max-Planck-Institut f\"{u}r Physik komplexer Systeme,
Bayreuther Str. 40,
Haus 16, 01187 Dresden, Germany.}
\author{C. A. Balseiro}
\address{Centro At\'{o}mico Bariloche and Instituto Balseiro, 8400 S.
C. de Bariloche, Argentina.}
\date{\today}
\maketitle

\begin{abstract}
    Using numerical techniques we study the zero temperature
properties of a substitutional magnetic impurity in a one-dimensional
correlated system described by the Hubbard model with repulsive
interaction $U$. We find that, while the static spin correlation maintains
a $2k_F$ periodicity, the charge correlations show Friedel oscillations with
a $4k_F$ periodicity for large values of $U$. The static on-site
susceptibility decreases as the interactions in the chain are turned on
 and we interpret this as an increase of the Kondo temperature.
For large values of $U$ the short range antiferromagnetic correlations
dominate over the collective Kondo singlet formation.
\end{abstract}

\narrowtext

    Recently much effort has been devoted to the understanding
of the effect of impurities in strongly correlated systems
\cite{and1,poil,fulde}. However,
while the problem of impurities in non-interacting systems like the
Kondo \cite{kondo,natan,wilson}, Anderson \cite{and2,wieg} and Wolff
\cite{wolff} models is well understood, very little is known about the
effect that the band correlations may have on the thermodynamic and
ground state properties of impurity systems.
Some experiments on high-Tc materials \cite{brugger,hardy,ishida} show
the effect of impurities interacting with a strongly correlated
$Cu-O$ plane. For instance,  heavy fermion behaviour has been observed in
$Nd_{1.8}Ce_{0.2}CuO_2$ below $T=0.3K$ \cite{brugger}.  This behaviour
has been modeled by considering a lattice of 4f $Nd$ ions interacting with a
system of strongly correlated conduction electrons. The strong
antiferromagnetic correlations in the
$Cu-O$ plane is taken into account
by breaking the spin symmetry \cite{fulde}.
Recent theoretical descriptions of the Kondo effect in a Luttinger liquid
have been made \cite{lee,furu} finding an algebraic scaling of the Kondo
temperature ($T_K$) with the Kondo exchange coupling
instead of the exponential
relation for the conventional Kondo problem in Fermi liquids.

In this paper we perform a finite-size calculation of a single impurity
embedded in a correlated electron system.
We study the case of a substitutional impurity in a Hubbard ring. We
consider a magnetic impurity and present results for the ground state
properties such as charge and spin correlation functions as well as
dynamical properties: local density of states at the impurity site and
frequency dependent spin susceptibility.

The model Hamiltonian reads:

\begin{equation}
\label{ham}
H=-\sum_{i,\sigma}t_i(c^{\dagger}_{i+1,\sigma}c_{i,\sigma} + hc)+
U\sum_{i\neq l}n_{i\uparrow}n_{i\downarrow} +
U_ln_{l\uparrow}n_{l\downarrow} +
\epsilon_l(n_{l\uparrow}+n_{l\downarrow})
\end{equation}
Here $U_l$ is the repulsion on the impurity site $l$, $\epsilon_l$
its diagonal energy, $t_{l-1}=t_l=t'$ and $t_i=t$ for all the
other sites. Nevertheless we will
consider $t'=t=1$ because we find that the results are qualitatively
the same even if $t'\ll t$.
As we are interested in a magnetic impurity we
take $\epsilon_l\ll -t$ and $\epsilon_l+U_l \gg t$.

    We use the Lancz\"os technique to obtain the ground state of
Hamiltonian (\ref{ham}) on an $N$-site chain with $N_e$ electrons.
 In order to get a non-degenerate
ground state we consider periodic
(antiperiodic) boundary conditions for $N_e/2$
odd (even) (closed shell condition).

    In the one-dimensional Hubbard model (in our case it corresponds to
$U_l=\epsilon_l=0$ and $t'=t$) the asymptotic charge and spin
correlation functions are given by
\cite{schulz}:
\begin{eqnarray}
\langle n(x)n(0) \rangle &=& K_{\rho}/(\pi x)^2 + A_1 \cos
(2k_F x)x^{-1-K_{\rho}} \ln^{-3/2}(x) + A_2 \cos (4k_F
x)x^{-4K_{\rho}} \\
\langle {\bf S}(x)\cdot {\bf S}(0) \rangle &=& 1/(\pi x)^2 +
B_1 \cos(2k_F x) x^{-1-K_{\rho}} \ln^{1/2}(x)
\end{eqnarray}
with interaction-dependent parameters$A_1$, $A_2$, $B_1$ and
$K_{\rho}$.
For large $U$ it is expected that the $4k_F$ term of the
charge correlation function is dominant \cite{and3}. This is confirmed
by  numerical results done in finite systems \cite{sorella}.
As we show below, this behaviour in the correlation functions
dominates the charge and spin screening of an impurity embedded in
a one-dimensional system.

    In Fig. \ref{fig:one} we show results for the static spin
$(\langle S^z_l S^z_j \rangle)$ and charge $(\langle n_l n_j
\rangle)$ correlation functions for the quarter filled case for
different values of $U$. We see that the interactions in the chain
enhance the antiferromagnetic correlations and the periodicity remains
$2k_F$ for the spin correlation. On the other hand,
for the charge correlations we find that, as
$U$ increases, the amplitude of the oscillations decreases and the main
periodicity changes from $2k_F$ to $4k_F$ at $U\simeq 4t$.
The behaviour of the system at large $U$ can be understood
from the $U=\infty$ limit were the charge dynamics can be described by
non-interacting spinless fermions with $k_F$ replaced by $2k_F$.
Consequently one expects a contribution proportional to $\cos(4k_F)$
in the density-density correlation functions in this limit.
Note that $\langle
n^2_l \rangle \simeq 1$ and $\langle (S^z_l)^2 \rangle \simeq 1/4$ for
all $U$ as corresponds to a magnetic impurity with $\epsilon_l \ll -t$ and
$\epsilon_l+U_l \gg t$.

    We also calculate numerically \cite{carlos} the density of
states at the impurity site $\rho_l(\omega)$. In Fig. \ref{fig:two} we
present results for the doped case where the delta peaks have been
broadened by a Lorentzian only for visualizing purposes. Important
peaks exist close to $\epsilon_l$ and  $\epsilon_l+U_l$ (the latter
ones are out of the range of the figure).
The lower and upper Hubbard bands can clearly
be recognized for $U=6t$ (Fig. 2c).
A large density of states near the Fermi energy can be
seen for all $U$ values.
 This enhancement is due to the scattering potential
introduced by the impurity. Due to the finiteness of the system, we can
see two enhanced peaks, one beneath and one above the Fermi surface but these
should become one resonant peak in the infinite system. In the Anderson
model (in the spin fluctuating limit) this corresponds to the
Kondo resonant peak and its width is proportional to $k_BT_K$.
Because we have a finite system
it is impossible to extract from these figures the width of
the resonant peak. However, we observe that as $U$ increases, the
most prominent peaks separate somewhat; this could suggest that
the resonant peak gets wider in the thermodynamic limit, implying
an increase of the Kondo temperature.
 For the half filled case the resonant peak at $U=0$
disappears at finite $U$ due to the formation of the Mott-Hubbard gap.
The results presented above are consistent with the dependence of the
static spin susceptibility with the interactions of the band presented below.

The local frequency and temperature-dependent spin susceptibility is given by
the following expression:
\begin{equation}
\chi'_l(\omega)=\frac{1}{\pi} \int_{-\infty}^{\infty}
\tanh\left ({\omega^{\prime}\over
2k_BT}\right ) {Im \langle\langle S^+_l,S^-_l \rangle\rangle
^R_{\omega^{\prime}+i\delta}
\over \omega-\omega^{\prime}+i\delta} d\omega^{\prime}
\end{equation}
where $\langle\langle \;\;\rangle\rangle^R$ is the retarded Green
function and $\chi'_l$ is in units of $(g\mu_B/2)^2$.

The static and zero temperature limit is:

\begin{equation}
\label{chi}
\chi'_l(\omega=0)=-2 Re \left.{\langle \psi_0 | S^+_l {1 \over
\omega+E_0-H+i\delta} S^-_l |\psi_0 \rangle } \right |_{\omega=0}=
2 \sum_\nu {|\langle\nu |S^-_l|\psi_0 \rangle |^2 \over E_\nu -E_0}
\end{equation}
where $|\psi_0 \rangle$ is the ground state of Hamiltonian (\ref{ham})
with energy $E_0$ and $|\nu \rangle$ and $E_\nu$ are the excited states and
their energies, respectively.
This static susceptibility  describes essentially the inverse of the
difference in energy between the first excited and the ground state
$(\Delta E)$. This can be seen by looking at the imaginary part of the
response function:
\begin{equation}
\label{imag}
\chi''_l({\omega})=-\frac{1}{\pi} Im \langle \psi_0 |S^+_l{1\over
\omega+E_0-H+i\delta} S^-_l|\psi_0 \rangle
\end{equation}
This is shown in Fig. 3 where we can see that
the first excited state has much more weight than the
other states of higher energy and is clearly distinguished from them.
In the half filled case, instead, the spin excitation spectrum is more
complicated but still a clear first excitation peak can
be distinguished for small $U$.
We have also checked the spin of the ground state and first excited
state, finding they are a singlet and triplet respectively.

    For $U=0$, {\it i.e.} no correlations in the band,
Hamiltonian (\ref{ham}) corresponds to the Wolff model in 1D \cite{wolff}. In
this case it is known that the properties of the system are very
similar to those of the Anderson model in the spin fluctuating limit
\cite{schlo}.
In particular, as calculated in \cite{zlatic} and checked numerically
by us, $\chi'_l$ increases with the impurity interaction $U_l$.

    In Fig. \ref{fig:chi} we plot the susceptibility as a function
of $U$ for several fillings. The case  $N_e=2$ (not shown)
 is special because
one electron is localized at the impurity site so there are no
many-body effects on the band.
 Provided $\epsilon_l\ll -t$ and $\epsilon_l+U_l\gg
t$ (as is our case),
 a simple second-order perturbation calculation shows that $\Delta E$
always decreases with $U$.
When more electrons are added to the system and the correlations start
playing an important role, we find that $\chi'_l$ has a non-monotonic
behaviour as a function of $U$. For small $U$ the susceptibility
decreases, it reaches a minimum and then starts to increase. This behaviour can
also be seen from (\ref{imag}) where the first peak goes towards
higher energies for small $U$ but then tends to zero (very high
$\chi'_l$) for large $U$ (see Fig. \ref{fig:im}). The susceptibility
for the half filled case, instead, always increases with $U$.
The inset of Fig.  \ref{fig:chi} shows the susceptibility as a function of
$U$ for a quarter filled band for $N=8$ and $N=12$. Here
we see that the lowering of $\chi'_l$ for low $U$
is more pronounced for larger systems suggesting that it will be an
important feature in the thermodynamic limit.

    In conclusion, we have calculated numerically some  properties
of dilute magnetic impurities embedded in a one-dimensional correlated
electronic system. Although we have performed our calculations in a
finite system where the low energy characteristics may be missed, this
method gives an insight on the properties of the system at zero
temperature.

    The static correlation functions reflect the non-Fermi liquid
behaviour of correlated electrons in one dimension: While the
periodicity of the spin-spin correlation remains $2k_F$, the Friedel
oscillations in the density-density correlation have a $4k_F$
periodicity at large $U$. This could probably be measured
experimentally with X-ray scattering in doped quasi-one-dimensional
systems with dilute impurities.

    We also find a large local density of states
near the Fermi energy due to the impurity scattering potential. The
separation of the most important peaks may indicate that the width of
the resonant peak in the thermodynamic limit increases.

    The susceptibility, which measures essentially the inverse of
the energy difference between singlet ground state and first excited
triplet state (interpreted here as proportional to $T_K$),
 shows a non-monotonic behaviour for the
doped case. For $U=0$ a bound singlet state is formed around the
impurity \cite{schlo}. In this non-interacting limit only a small
fraction of electrons (those within $k_BT_K$ of the Fermi surface)
are involved in the extended
singlet state. When the interactions in the band
are turned on, the effective mass of the carriers increases, enlarging
the density of states near the Fermi energy. As more electrons are involved
in the screening of the impurity, the spin compensating cloud becomes more
localized and as a consequence, $T_K$ is larger
{\it i.e.} $\chi'_l$ decreases. For large $U$ this picture
holds no longer because the strong antiferromagnetic correlations
inhibit the formation of a collective singlet state around the
impurity and the susceptibility increases monotonously.
In the half filled case a gap opens for finite $U$ and there is no
formation of a collective singlet state around the impurity site and
the susceptibility also increases.
 In this case one expects
the susceptibility to diverge at large $U$ (when $U_l \to \infty$)
because conformal field theory
 studies \cite{eggert} show that the fixed point of a substitutional
spin with antiferromagnetic coupling in a Heisenberg chain is a
`healed' Heisenberg chain (with $J=2t^2/U$).

For the highly doped case, our system is in the Luttinger liquid regime
and our results for $T_K$ at small $U$
are qualitatively similar to those of Ref.~\cite{furu}
(small and positive $g_2$ in their case and  small
Kondo interaction parameter $J$). We find that in the Hubbard model
the antiferromagnetic correlations play an important role and
compete with the collective Kondo singlet formation, dominating at
large values of $U$.

It is interesting to study whether in higher dimensions the features
encountered here, {\it i.e.}the the non-monotonous behaviour
of the Kondo temperature as a consequence of a competing interaction
between Kondo singlet formation and antiferromagnetic interactions,
persist. Of course, other methods than exact diagonalization techniques
are required. The doubling of the periodicity of the Friedel oscilations in
the density-density correlation function is,
however, a characteristic of Luttinger liquids.

{}~\\
{}~\\

We would like to thank P. Horsch, H. Eskes, A. Ruckenstein and
E. M\"uller-Hartmann for  profitable discussions and especially
P. Fulde for valuable discussions and  hospitality.

\begin{figure}
\caption{Static spin-spin (a) and density-density (b) correlation functions
vs. the distance to the impurity site
for $N=12$, $Ne=6$ ($k_F=\pi/4$), $U_l=100t$ ,$\epsilon_l=-10t$ and
$U=0$ (circles), $U=t$ (squares), $U=2t$ (diamonds), $U=4t$ (triangles)
and $U=10t$ (plus).}
\label{fig:one}
\end{figure}
\begin{figure}
\caption{Density of states at the impurity site $l$ for
$N=10$, $Ne=6$, $U_l=100t$ and $\epsilon_l=-10t$ for
$U=0$ (a), $2t$ (b)
and $6t$ (c). Full (dotted) lines: occupied (empty) states. }
\label{fig:two}
\end{figure}
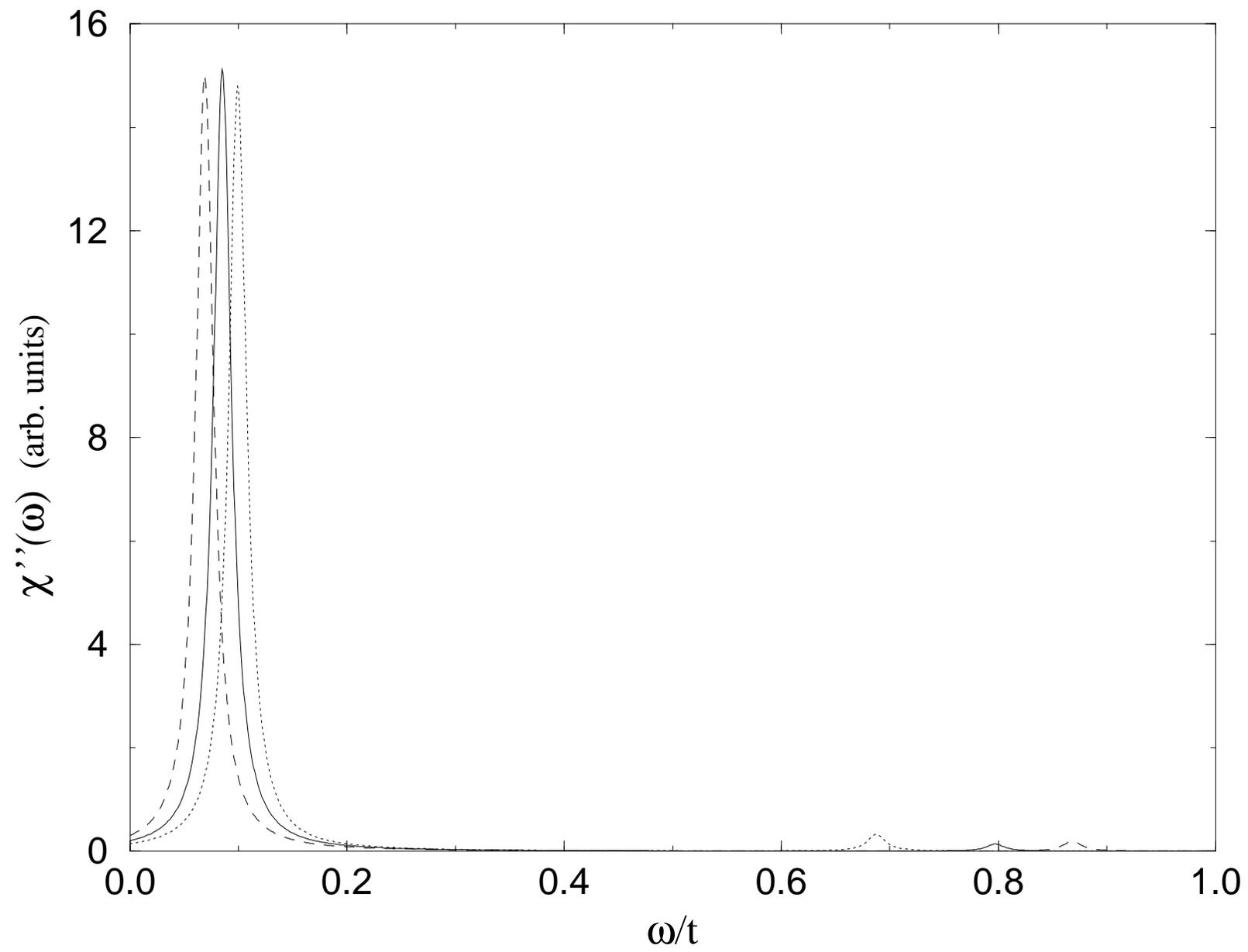
\begin{figure}
\caption{$\chi''_l(\omega )$ for the same set of parameters as in Fig. 1 but
for $U=0$ (full line), $U=4t$ (dotted line) and $U=30t$ (dashed line)}
\label{fig:im}
\end{figure}
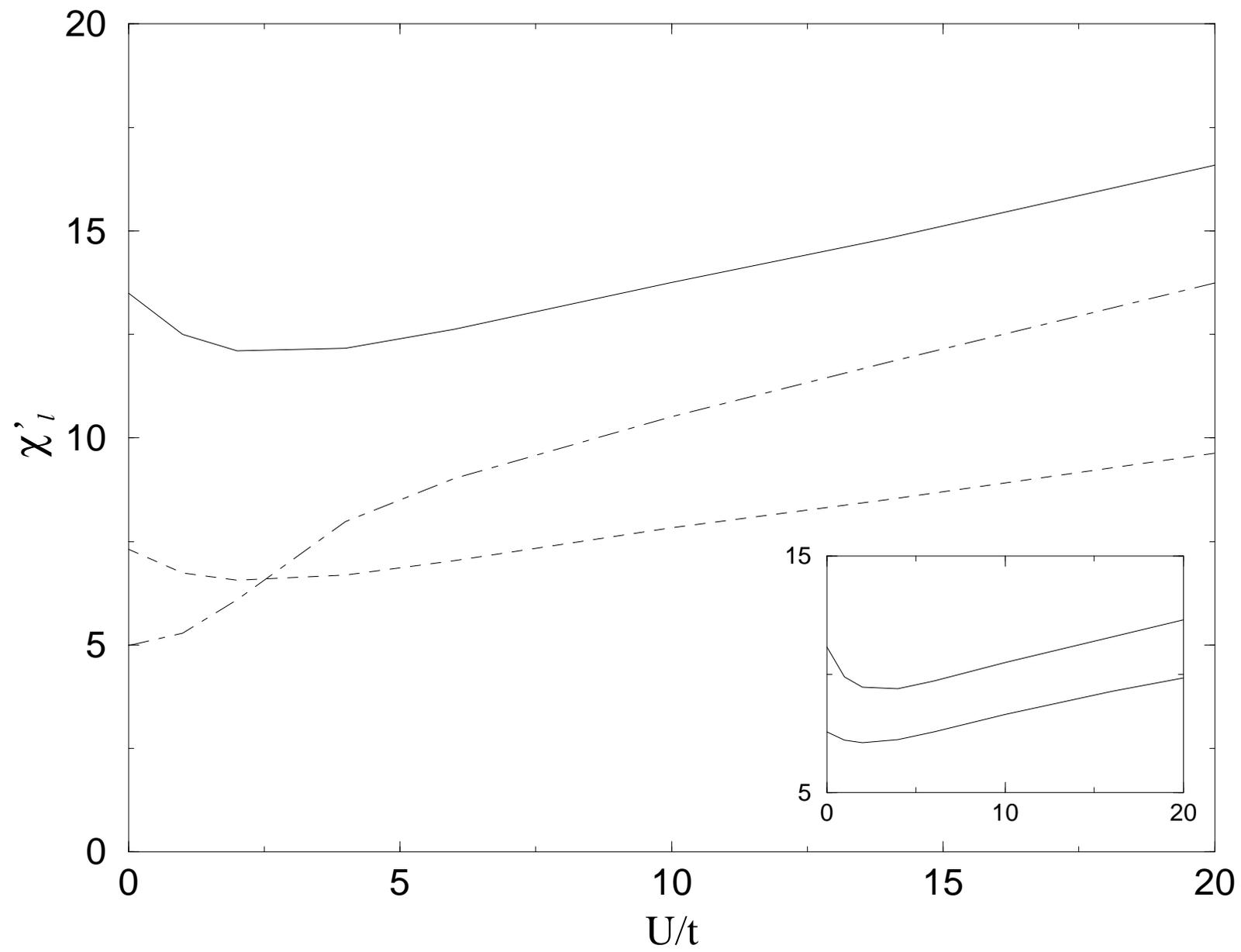
\begin{figure}
\caption{Local static susceptibility $\chi'_l$ as a function of $U$ for
$U_l=100t$, $\epsilon_l=-10t$,
$N=10$ and different fillings: $Ne=4$ (full line), $6$ (dashed line)
and $10$ (dashed-dotted). The inset shows $\chi'_l$ as a function
of $U$ for a quarter filled band for $N=8$ (lower curve) and
$N=12$ (upper curve), for $U_l=20t$ and $\epsilon_l=-10t$. }
\label{fig:chi}
\end{figure}
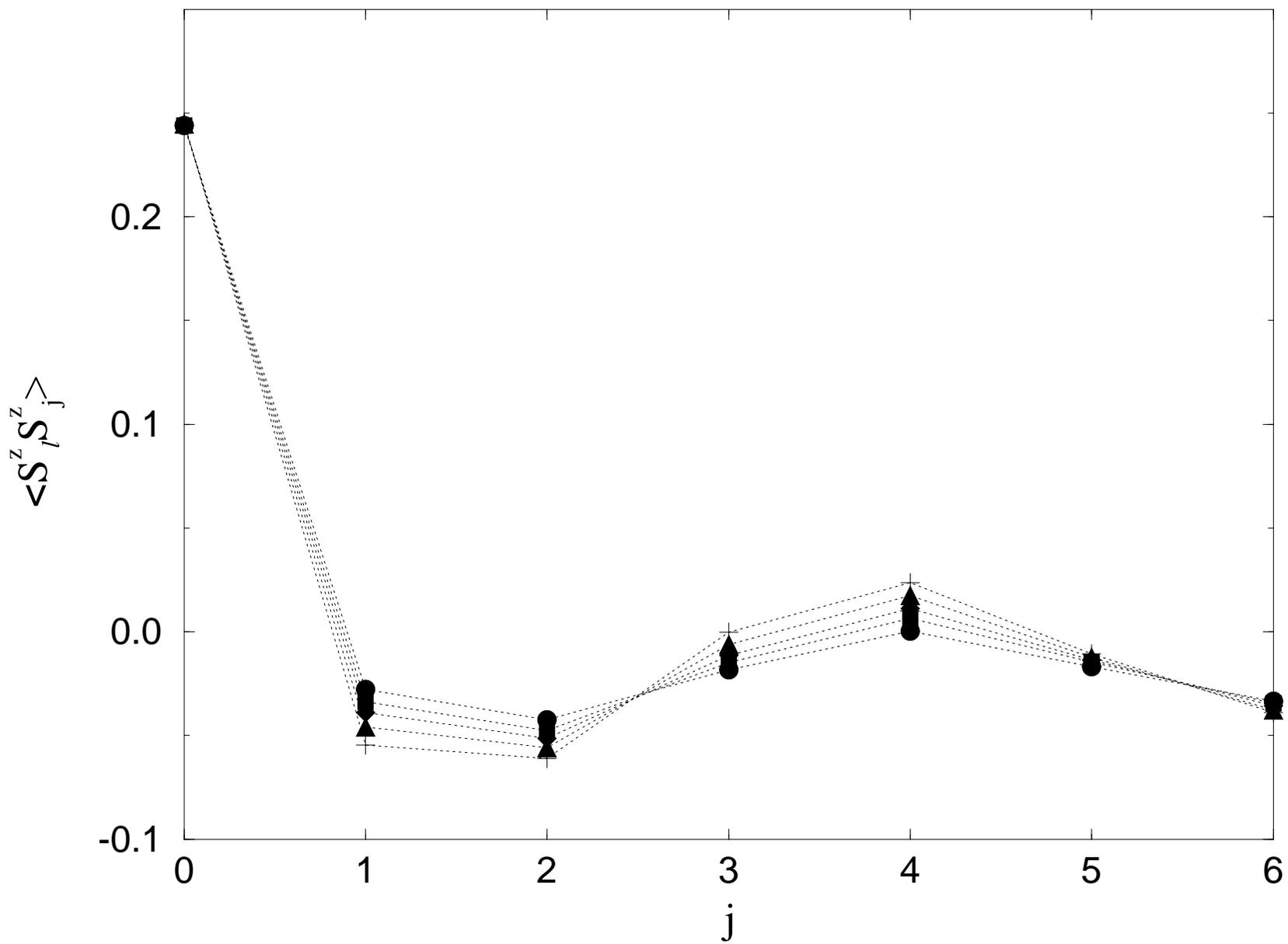
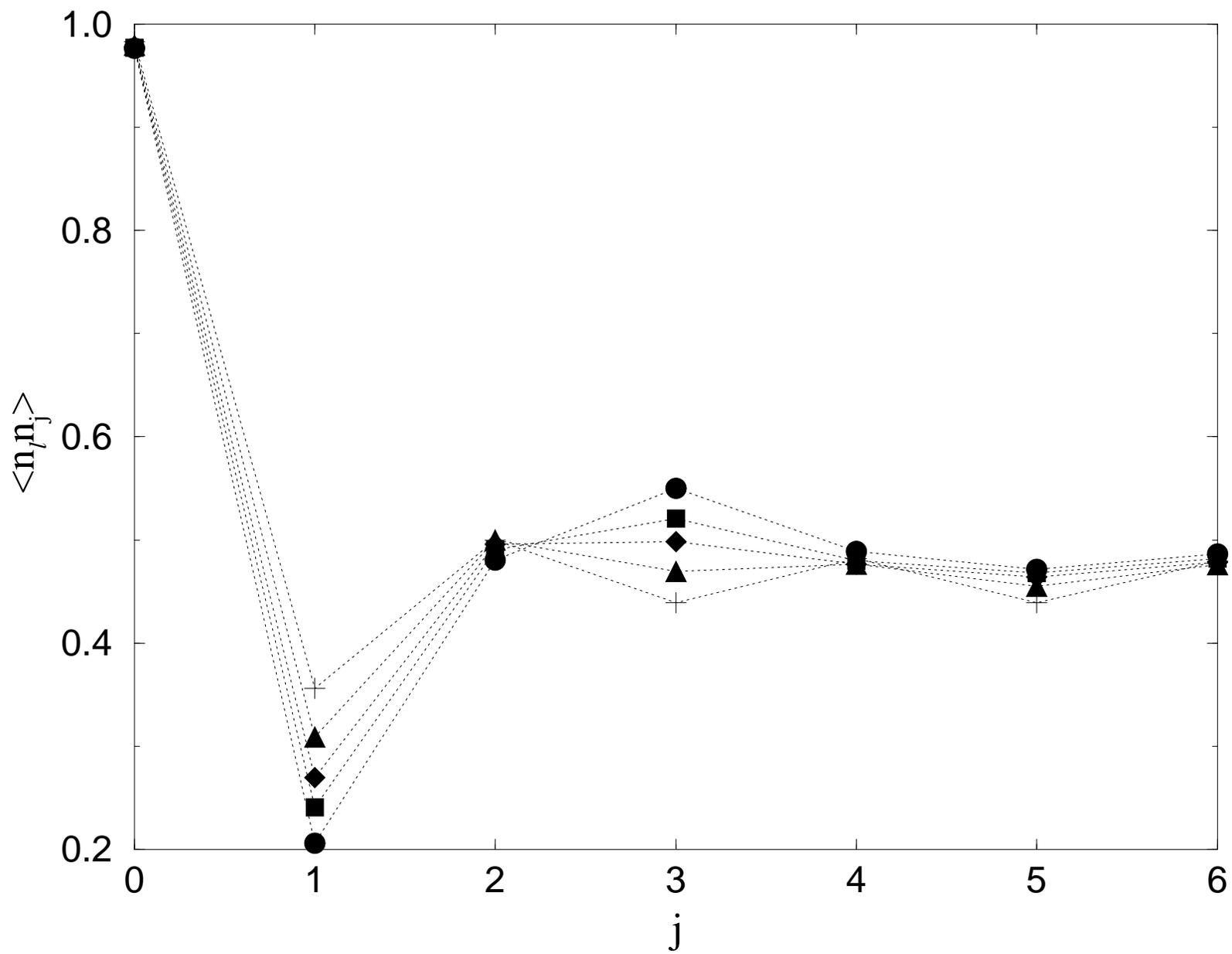
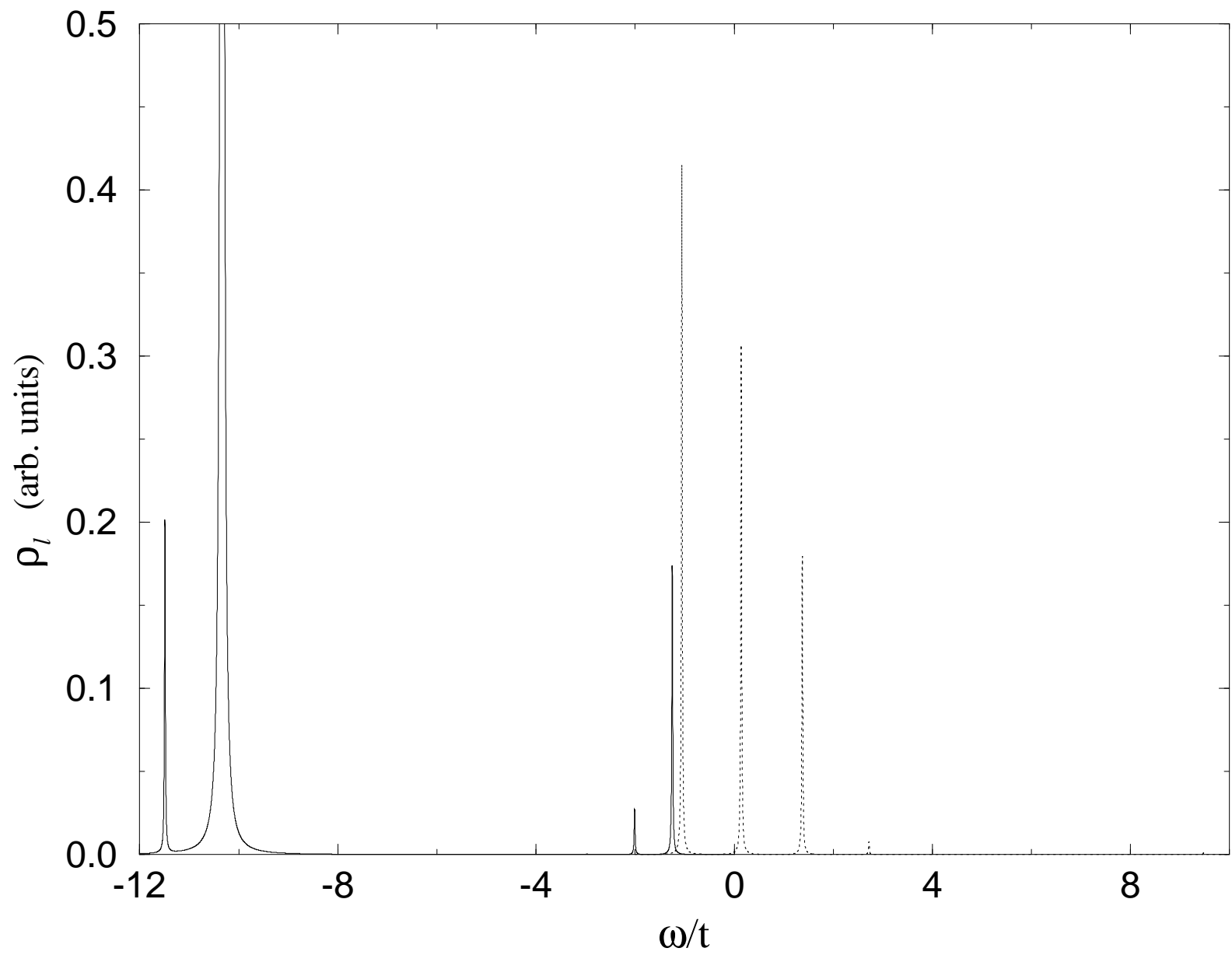
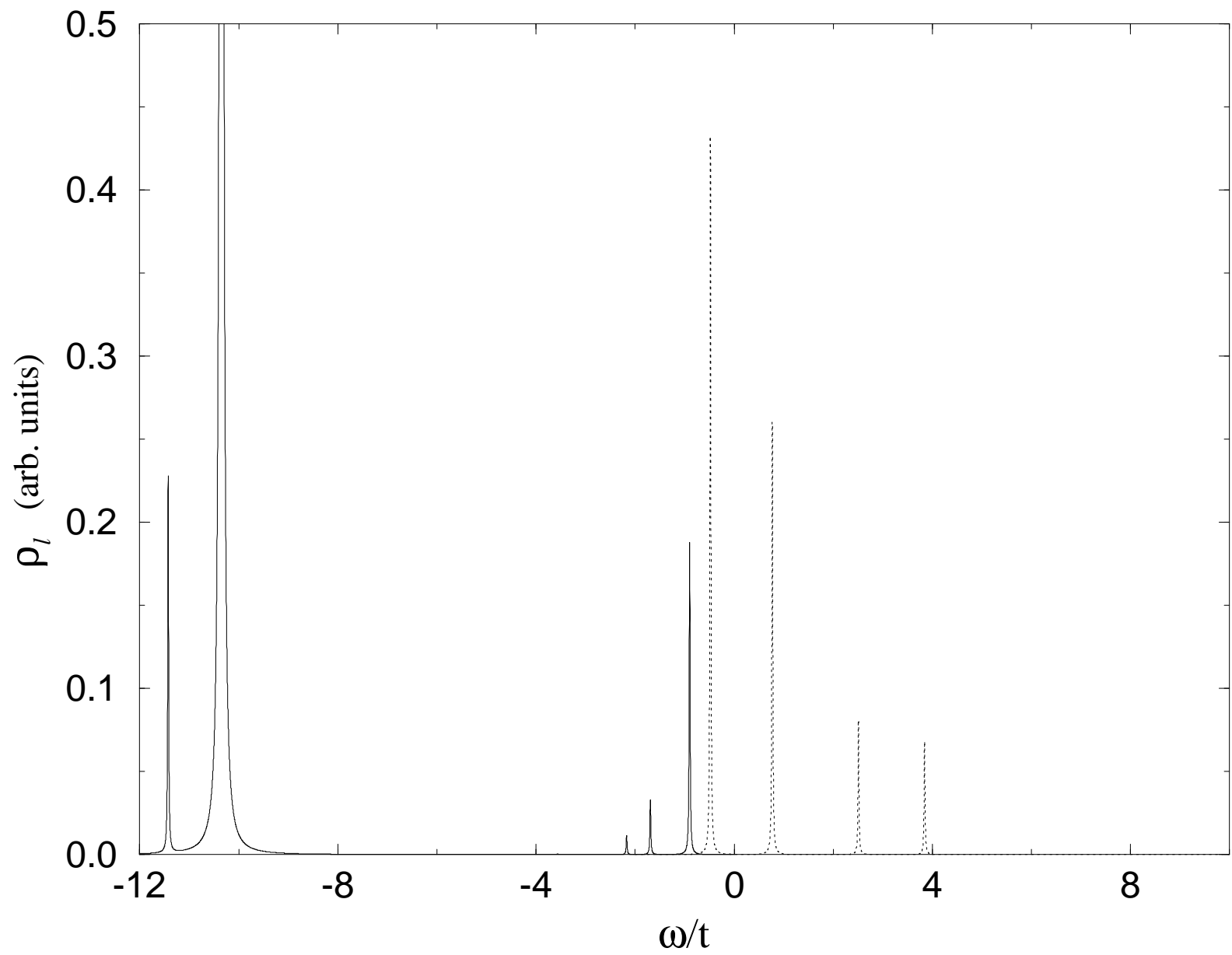
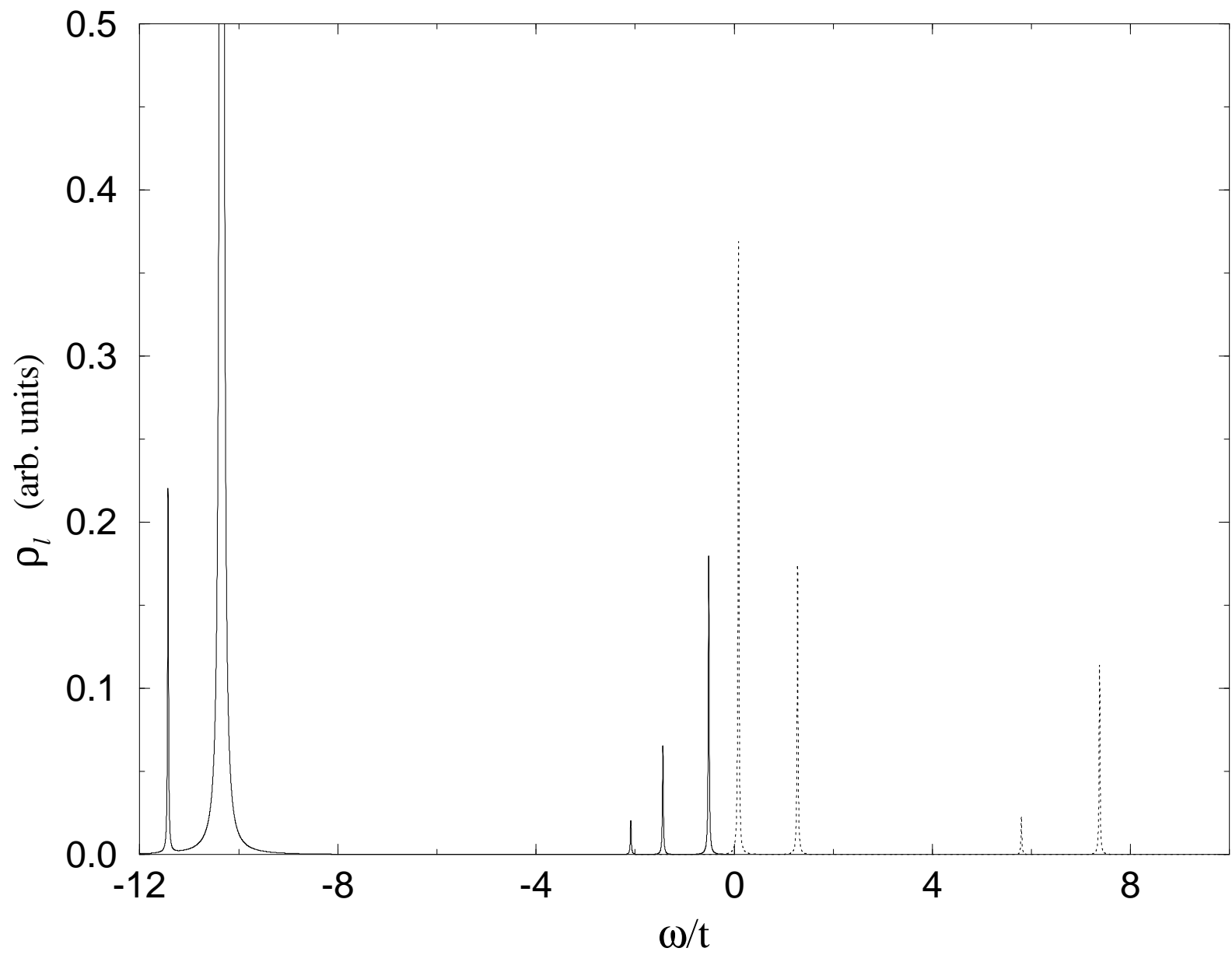
\end{document}